\documentclass[useAMS,usenatbib,usegraphicx]{mn2e}
\usepackage{times}
\usepackage{multicol}

\newcommand{\AIPS}{{$\cal AIPS\/$}}

\def\mic{$\mu$m}
\def\Msun{M$_{\odot}$}
\def\gs{\mathrel{\raise0.35ex\hbox{$\scriptstyle >$}\kern-0.6em
\lower0.40ex\hbox{{$\scriptstyle \sim$}}}}
\def\ls{\mathrel{\raise0.35ex\hbox{$\scriptstyle <$}\kern-0.6em
\lower0.40ex\hbox{{$\scriptstyle \sim$}}}}

\title[Submillimetre polarimetry of Cas A]
      {Cassiopeia A: dust factory revealed via submillimetre polarimetry}
\author[L.\ Dunne et al.]
       {L.\ Dunne,$\!^1$ 
        S.\,J.\ Maddox,$\!^1$ 
        R.\,J.\ Ivison,$\!^{2,3}$ 
        L.\ Rudnick,$\!^4$
        T.\,A.\ DeLaney,$\!^5$       
	B.\,C.\ Matthews,$\!^6$
        \and
        C.\,M.\ Crowe,$\!^1$  
        H.\,L.\ Gomez,$\!^7$
        S.\,A.\ Eales$\!^7$ and
        S.\ Dye$\!^7$                
        \vspace*{1mm}\\
        $^1$ School of Physics \& Astronomy, University of Nottingham,
             University Park, Nottingham NG7 2RD\\
        $^2$ UK Astronomy Technology Centre, Royal Observatory, Blackford
             Hill, Edinburgh EH9 3HJ\\
        $^3$ Institute for Astronomy, University of Edinburgh, Blackford Hill,
             Edinburgh EH9 3HJ\\
        $^4$ Department of Astronomy, University of
             Minnesota, 116 Church Street, SE Minneapolis, MN 55455, USA\\ 
        $^5$ MIT Kavli Institute, 77 Massachusetts Avenue, Room NE80-6079,
             Cambridge, MA 02139, USA\\
        $^6$ Herzberg Institute of Astrophysics, National Research Council
             of Canada, 5071 West Saanich Road, Victoria, BC, V9E 2E7, Canada\\
        $^7$ Department of Physics \& Astronomy, Cardiff University,
             Queen's Buildings, The Parade, Cardiff CF24 3AA
}

\pagerange{000--000}

\begin{document}

\maketitle

\begin{abstract}
If Type-{\sc ii} supernovae -- the evolutionary end points of
short-lived, massive stars -- produce a significant quantity of dust
($>$0.1\,\Msun) then they can explain the rest-frame far-infrared
emission seen in galaxies and quasars in the first Gyr of the
Universe. Submillimetre (submm) observations of the Galactic supernova
remnant, Cas~A, provided the first observational evidence for the
formation of significant quantities of dust in Type-{\sc ii}
supernovae. In this paper we present new data which show that the
submm emission from Cas~A is polarised at a level significantly higher
than that of its synchrotron emission. The orientation is consistent
with that of the magnetic field in Cas~A, implying that the polarised
submm emission is associated with the remnant. No known mechanism
would vary the synchrotron polarisation in this way and so we
attribute the excess polarised submm flux to cold dust within the
remnant, providing fresh evidence that cosmic dust can form
rapidly. This is supported by the presence of both polarised and
unpolarised dust emission in the north of the remnant, where there is
no contamination from foreground molecular clouds. The inferred dust
polarisation fraction is unprecedented ($f_{\rm pol}\sim 30$ per cent)
which, coupled with the brief timescale available for grain alignment
($<$300\,yr), suggests that supernova dust differs from that seen in
other Galactic sources (where $f_{\rm pol}=2-7$ per cent), or that a
highly efficient grain alignment process must operate in the
environment of a supernova remnant.
\end{abstract}


\section{Introduction}
\label{sec:intro}

The large quantities of dust seen in high-redshift quasars and
galaxies \citep{priddey03, priddey08, wang07, wang08}, at a time when
the Universe was only $\sim$1\,Gyr old, suggests that a rapid
mechanism for dust production must operate. Type-{\sc ii} supernovae
(SNe) are good candidates for these dust factories as they evolve to a
dust-producing phase in only a few hundred Myr and contain a high
abundance of heavy elements. Theory predicts that each Type-{\sc ii}
SNe should produce $\sim$0.1--1\,\Msun\ of dust \citep{tf01, nozawa03,
bs07} and, if true, this can account for the dust observed at high
redshift \citep{me03, dwek07}.

Evidence for this quantity of dust forming in young SNe in the local
Universe has been scant and controversial. Observations from
mid-/far-infrared (IR) satellites ({\em IRAS\/}, {\em ISO\/} and {\em
Spitzer\/}) detect only warm dust and find orders of magnitude less
than predicted by theory \citep{dwek87, lagage96, douvion01, blair07,
borkowski06, williams06, sugerman06, meikle07}, e.g.\ the maximum warm
dust mass inferred for Cas~A from {\em Spitzer\/} data is
$0.03-0.05$\,\Msun\ \citep{rho08}. Submillimetre (submm) observations
of Cas~A seemed to provide the first evidence for large ($\sim$2\,\Msun)
quantities of colder dust manufactured in the supernova explosion
\citep[][hereafter D03]{dunne03}\defcitealias{dunne03}{D03}. However,
subsequent observations of CO and OH towards Cas~A suggested that some
or all of the submm emission may originate from dust in a foreground
molecular cloud complex rather than the remnant \citep{krause04,
wb05}. Although the level of foreground contamination is highly
uncertain -- because of the difficulties in converting a molecular
line intensity into a submm flux density -- the molecular data have
cast doubt on the idea that significant quantities of dust can form
rapidly in SNe.

Another explanation for the submm excess, this time without a
significant mass of dusty material, was proposed by \citet{dwek04}.
Elongated iron needles with a much higher submm emissivity than
canonical dust can also account for the far-IR and submm spectral
energy distribution (SED) of Cas~A. If present in the general ISM,
they should contaminate the polarisation signal from the CMB which
could have importance consequences \citep[e.g.][]{bowden04}.

Polarimetry at 850\,\mic\ is extremely challenging but provides one way
to test the competing hypotheses for the submm emission in Cas~A: cold
dust in Cas~A, foreground contamination, or iron needles. We report
such observations here, with the paper organised as follows: \S2
describes the submm and radio polarimetry; \S3 presents the results
and investigates the robustness of the submm polarimetry; in \S4, we
use the polarimetry to place constraints on the fraction of submm
emission arising from the remnant and comment on the implications for
grain alignment theories.

\section{Observations and data reduction}
\label{sec:obs}

\subsection{Submm observations}
\label{submmobs}

\begin{figure}
\centering
\includegraphics[width=8cm]{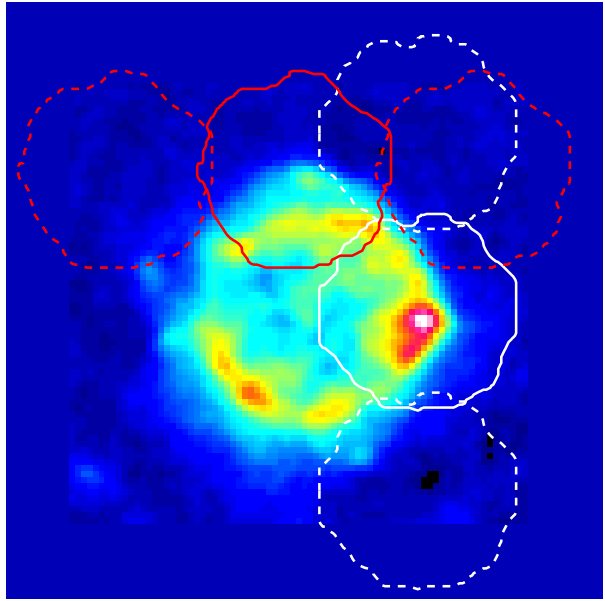}
\caption{Cas~A at 850\,\mic, from \citetalias{dunne03}, showing the
regions observed with the SCUBA polarimeter. On-source fields are
denoted by solid lines while chop positions are shown by dashed
lines. Position W (white) is the brightest region at radio and submm
wavelengths and is coincident with a CO emission-line peak. Position N
(red) is where the dust-to-synchrotron ratio is lowest and where there
is no foreground emission.}
\label{fig:regionsF} 
\end{figure}

Our submm observations took place during 2004 October 17--18 using the
SCUBA polarimeter \citep{murray97, holland99, greaves03} at the 15-m
James Clerk Maxwell Telescope (JCMT), Hawaii\footnote{The JCMT is
operated by the Joint Astronomy Centre on behalf of the Science and
Technology Facilities Council of the United Kingdom, the Netherlands
Organisation for Scientific Research, and the National Research
Council of Canada.}. We used SCUBA at 850\,\mic\ where the array
contains 37 bolometers. The instrument comprises a rotating quartz
half-wave plate and a fixed photo-lithographic-grid analyser in a
module that is placed in front of the SCUBA cryostat's entrance
window. The weather conditions were exceptional -- very stable with
very little sky noise, with $\tau_{\rm 230GHz}\ls 0.04$ throughout.
Two positions on the bright rim of Cas~A were observed
(Fig.~\ref{fig:regionsF}), one in the west (hereafter W) and one in
the north (hereafter N). Our original Cas~A submm map
\citepalias{dunne03} was taken in SCUBA's `scan-map' mode; for the
polarimetry we were forced to make a `jiggle' map, stepping the
secondary mirror of the telescope in a 16-point pattern in order to
provide a fully sampled image at 850\,\mic, whilst chopping a distance
of 180\,arcsec at around 7\,Hz (the maximum possible displacement) --
to remove the atmospheric signal, and nodding the telescope every
32\,s to correct for slowly varying sky gradients. The chop positions
were chosen to avoid the remnant as much as possible;
Fig.~\ref{fig:regionsF} shows the location of the fields and their
chop positions. A detailed discussion of the potential effects of
chopping onto emission in the reference positions is presented in
\S3. A full 16-point jiggle was taken at each of 16 positions of the
half-wave plate, separated by 22.5$^{\circ}$. Thus 16 maps complete
one full rotation, yielding four redundant polarisation
observations. We co-added five and three such observations for the
Western and Northern regions, respectively yielding total
integration times of 65 mins for the Western region and 45 mins for
the Northern region.

The submm data were reduced using the SCUBA User Reduction Facility
\citep[SURF --][]{jl98} and {\sc polpack} \citep{bg01}. The basic
analysis consisted of removing the nod, flatfielding, correcting for
extinction, flagging bad bolometers, removing sky noise and clipping
noisy data. For the extinction correction, we employed a polynomial
fit to the Caltech Submillimeter Observatory's 230-GHz skydip data. To
remove any residual time-varying sky level, we made preliminary maps
in order to choose bolometers lying in emission-free regions and then
subtracted their median signal in every second of integration from the
full array. We experimented with different sky bolometers and
different thresholds for removing noisy bolometers to assess their
impact on the polarimetry measurements. We found some small changes in
the derived polarisation parameters but all maps were consistent
within the errors (see \S3 for more details).

Instrumental polarisation (IP) arises through the SCUBA optics and the
GORE-TEX membrane in front of the dish \citep{greaves03}. IP is
elevation-dependent and can be characterised by making observations of
unpolarised planets. The IP was removed using the SURF task, {\sc
remip}, which uses a look-up table of bolometer IPs and corrects for
the elevation at the time of the observation. The uncertainty in the
IP removal is $\pm$0.5 per cent \citep{matthews01}. Once treated in
this way, each integration was re-gridded separately using a Gaussian
weighting function with a scale of 7\,arcsec (approximately half the
{\sc fwhm} for SCUBA at 850\,\mic) onto 6.2-arcsec pixels (the step
size of the jiggle pattern), ready for the polarisation analysis. From
each complete polarimetry observation, four degenerate maps are made
for each of the four equivalent waveplate angles. The four maps at
each angle are stacked and the variance for each pixel calculated from
the scatter between the four maps. {\sc polpack} was then used to
derive the Stokes $U$, $Q$ and $I$ values for each pixel in each
complete polarimetry observation by fitting the following function to
the data \citep{sa99} \[ {I_k}^{\prime} =
\frac{t}{2}\left[I+\epsilon(Q\cos2\phi_k + U\sin2\phi_k)\right] \]
where ${I_k}^{\prime}$ is the expected intensity in image $k$,
$\phi_k$ is the position angle of the wave plate after correction for
the parallactic angle for image $k$, $t$ is the analyser transmission
factor and $\epsilon$ is the polarising efficiency factor. The error
in each Stokes parameter is derived from the input variances.

These $I$, $Q$ and $U$ images were then median stacked for repeat
observations and a final data cube was created. Polarisation
fractions, $p$, and position angles (PAs), $\theta$, were derived from
the Stokes parameters \citep{hildebrand00} after correcting for the
Ricean bias -- i.e.\ the increase in $p$ that results from
constraining $p$ to be positive even when $Q$ and $U$ are consistent
with $p=0$. Ricean bias is a significant effect for low
signal-to-noise measurements (SNR $\ls$ 2).

We define the normalised Stokes parameters and associated errors as

\[
u = U/I; \hspace{15mm} q=Q/I  
\]

\[
\sigma_q = \left(\frac{\sigma^2_I\ q^2 + \sigma^2_Q}{I^2}\right)^{1/2}; \hspace{3mm}
\sigma_u = \left(\frac{\sigma^2_I\ u^2 + \sigma^2_U}{I^2}\right)^{1/2}
\]

\noindent
Debiased $p$ values are calculated using
\begin{equation}
p = \frac{\sqrt{q^2+u^2 - \Delta}}{I} \times 100 \; {\rm per \; cent}
\end{equation}

\noindent
where the bias is given by

\begin{equation}
\Delta = \frac{q^2\ \sigma^2_q + u^2\ \sigma^2_u}{q^2 + u^2}
\end{equation}

\noindent
The error in $p$ is $\sigma_p =\sqrt{\Delta} \times 100$ per cent. The
polarisation PA -- the position of the electric field vector measured
relative to Celestial North -- is given by

\begin{equation}
\theta = 0.5 \; \arctan\left(\frac{U}{Q}\right)
\end{equation}

\noindent
The error on $\theta$ is given by $\sigma_{\theta} =
\frac{\sigma_p}{P} \times 28.6^\circ$.

The polarisation data were binned to a resolution of 18\,arcsec to
improve the SNR. Vectors were chosen for the catalogue if $\sigma_p <
4$ per cent, thereby removing unreliable vectors in regions of high
noise. We have also imposed a cut of $\rm SNR \geq 2$ in $p$ as vectors with
below this level have large errors in $\theta$ and the bias correction
we make is really only applicable at $\rm SNR\geq 2$
\citep{hildebrand00}. The SNR cut affects only 2/57 vectors and has no
consequences for our results (see \S~\ref{submmradio}). The absolute
accuracy in position angle is limited by the systematics of
removing the sky noise. This was found to be $\pm 6^{\circ}$.

As discussed by \citet{greaves03}, there is a minimum
believable polarisation percentage within the main beam of SCUBA
polarisation maps due to potential contamination of the polarisation signal
from the extended sidelobes of the JCMT beam. The value of
the minimum believable polarisation is given by the relation

\begin{equation}
p_{\rm crit} \ge 2 \ p_{\rm sl} \ P_{\rm sl} \ \frac{S_{\rm sl}}{S_{\rm mb}}
\end{equation}

\noindent where $p_{\rm sl}$ is the polarisation in the sidelobe
(i.e.\ off-centre position) and $P_{\rm sl}$ is the power in the
sidelobe relative to the main beam (measured from planetary
observations). The flux at the position of an off-centre source
relative to the flux at the map centre is given by $S_{\rm sl}/S_{\rm
mb}$.  We have used observations of Uranus, obtained at the time of
the Cas A observations, to assess the value of $p_{\rm crit}$ and
thereby measure the sidelobe polarisation for our science data.

The highest value of $p_{\rm crit}$ will be measured for the largest
ratio of $S_{\rm sl}/S_{\rm mb}$. This value arises for our
observations of the northern rim of Cas~A, in which the brightest peak
in the field of view does not lie at the array centre. For these
observations, we measure a flux ratio of 1.1 relative to the centre of
the array. The power in the beam within this annulus (49.0 $<
\sqrt{\Delta \alpha^2 + \Delta \delta^2} <$ 61.8\,arcsec, where
$\alpha$ and $\delta$ represent R.A.\ and Dec.) is 0.014 relative to
the main beam, and the IP measured within the same annulus is $9.2 \pm
4.6$ per cent. The highest value of $p_{\rm crit}$ is therefore $0.3
\pm 0.2$ per cent. We can therefore accept that all values of $p$ in
excess of 0.5 per cent arise from the remnant itself.  Since all
measured values exceed this threshold by a significant margin, it is
clear that none of the measurements are due to polarisation in the
sidelobe positions.

\subsection{Radio observations}

Cas~A was observed with the National Radio Astronomy Observatory's
(NRAO\footnote{NRAO is a facility of the National Science Foundation
operated under cooperative agreement by Associated Universities,
Inc.}) Very Large Array (VLA) in 2000--01 using all four
configurations from the most extended (A) to the most compact (D).
Data were taken at four frequencies, 4605, 4720, 4860 and 4995\,MHz,
as summarised in Table~\ref{radioobsT}.

Standard calibration procedures were employed, as outlined in the
\AIPS\ cookbook\footnote{\rm www.aips.nrao.edu/cook.html} using 3C\,48
to set the flux density scale, 3C\,138 to calibrate the polarisation
and J2355+498 as a local phase calibrator.  After initial calibration,
multiple passes of self-calibration were performed to improve the
antenna phase solutions.

The \AIPS\ maximum entropy deconvolution routine {\sc vtess}, which
maximises smoothness in an image, was used to restore the total
intensity images and the corresponding routine {\sc utess}, which
maximises emptiness in an image, was used to restore the Stokes $Q$
and $U$ images. An 18-arcsec beam ({\sc fwhm}) was used for
deconvolution, to match the resolution of the submm data.  The
standard primary beam attenuation correction was applied in the {\sc
vtess} and {\sc utess} routines. The total restored flux density was
788\,Jy. Noise-corrected, linearly polarised intensity, polarisation
percentage, and polarisation PA images were made from the Stokes $I$,
$Q$ and $U$ maps.  The total polarised flux density was 28.9\,Jy,
resulting in an average polarisation percentage of 3.7 per cent.

\begin{table}
\caption{\label{radioobsT} Summary of VLA observations.}
\begin{tabular}{cccc}
\hline
\multicolumn{1}{c}{Configuration}&\multicolumn{1}{c}{Date}&\multicolumn{1}{c}{Bandwidth}&\multicolumn{1}{c}{Duration}\\
\hline
A & 2000.9 & 6.25\,MHz & 12\,hr\\
B & 2001.2, 2001.3 & 12.5\,MHz & 11\,hr\\
C & 2000.3 & 25.0\,MHz & 6.0\,hr\\
D & 2000.7 & 25.0\,MHz & 4.5\,hr\\
\hline
\end{tabular}
\end{table}

To enable a comparison with the submm data, the radio $IQU$ datacube
was aligned with the submm pixels using the Kernel Application Package
(KAPPA) task {\sc wcsalign}. Polarisation vectors were then calculated
in the same way as for the submm data, but using bespoke {\sc matlab}
scripts. Radio vectors with $\sigma_p < 2.4$ were chosen, providing
complete coverage of the regions where we have submm vector
information.

\section{Results}
\label{sec:res}

\begin{figure*}
\includegraphics[width=8.5cm]{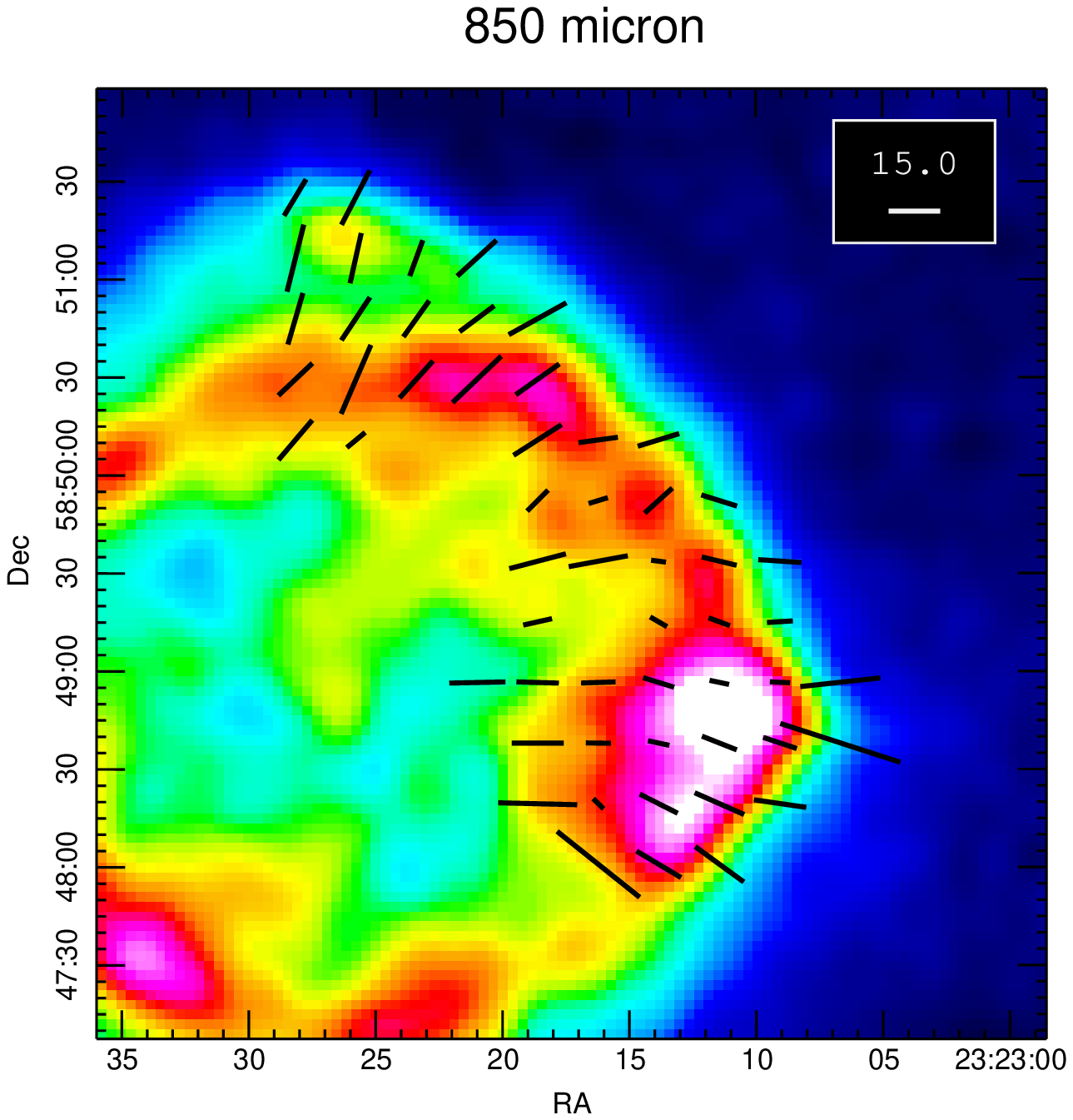}
\includegraphics[width=8.5cm]{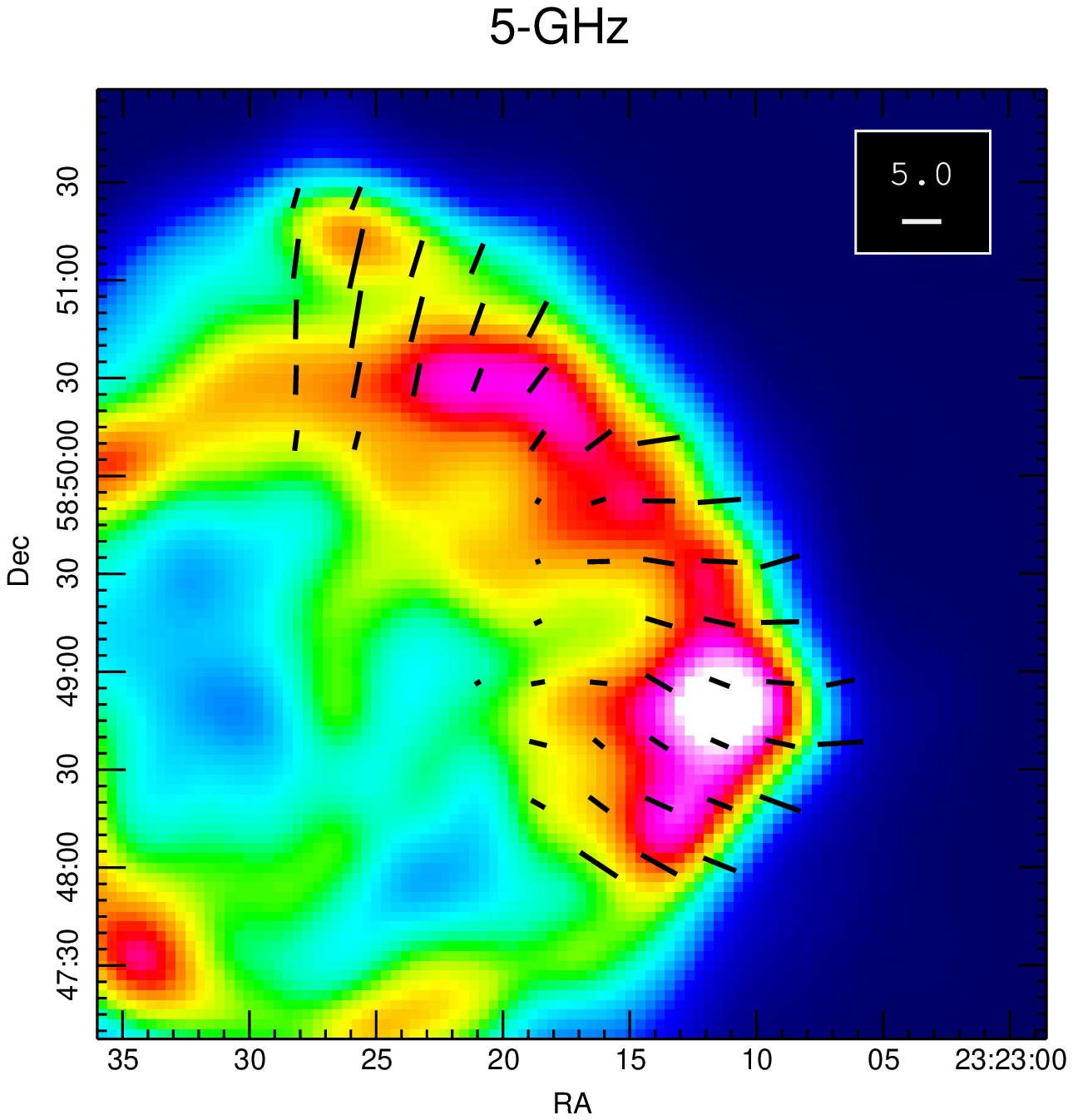}\\
\caption{{\em Left:} 850-\mic\ polarimetry vectors overlaid on the
850-\mic\ map from \citetalias{dunne03}. {\em Right:} 5-GHz
polarimetry vectors overlaid on the 5-GHz map. All maps are at the
same resolution, 18\,arcsec. Vectors are rotated by 90$^\circ$ to show
the direction of the magnetic field. The orientation of the submm
vectors clearly trace the approximately radial $B$ field of the
remnant. The 5-GHz vectors have been corrected for an average rotation
measure of $-110\,\rm{rad\,m^{-2}}$ .}
\label{mainpolF}
\end{figure*}

The SCUBA 850-\mic\ polarimetry vectors are shown in
Fig.~\ref{mainpolF}, alongside the 5-GHz radio vectors. The underlying
images are the 850-\mic\ scan-map from \citetalias{dunne03} and the
5-GHz image, both convolved to the same resolution as the binned
polarimetry data. The length of the vectors represents the degree of
polarisation, $p$, and their PA shows the direction of the magnetic
($B$) field (i.e.\ the $E$ vectors measured from the Stokes parameters
have been rotated by 90$^\circ$ for display purposes).

These images clearly show that the submm emission is significantly
more polarised than the radio emission, with vectors that trace the
approximately radial magnetic field in the remnant. Both traits
indicate a source of polarised emission in Cas~A above a general
extrapolation of the synchrotron polarised flux density.

We will now investigate in more detail the robustness of this result
and explore whether the submm polarisation is due to dust or
synchrotron emission.

\subsection{Robustness of results}
\label{sec:issues}

In this section we will describe our investigation into the robustness
of the data through comparing several reductions of the polarimetry
data and also through a simulation of the effects of chopping. We
conclude that the data are robust and that chopping cannot
conspire to affect our conclusions. Readers who wish to skip the
technical details can move on to \S~\ref{submmradio}.

\subsubsection{Data-reduction issues}

We explored the effect of using different sky bolometers and clipping
thresholds whilst reducing the submm data (\S\ref{sec:obs}). The
choice of sky bolometers is important as several effects can cause a
bolometer's signal to vary with time during a polarimetry observation:
\begin{enumerate}
\item{A changing sky level, which should correlate across all
bolometers.}
\item{Bolometers jiggling on and off a bright source. This is the
rationale behind choosing off-source bolometers to remove sky
variations across the whole array.}
\item{Bolometers may lie on regions of the same brightness throughout
an observation but see polarised emission which changes as a function
of wave-plate angle, and therefore as a function of time within the
observation.}
\end{enumerate}

If bolometers displaying effects (ii) or (iii) are used to remove sky
noise, they could introduce a temporal signal variation to the rest of
the array and cause systematic changes in the derived polarisation
properties.

We tried two sky-removal techniques: the first used as many bolometers
as possible, whilst avoiding bright emission regions; a second, more
conservative approach used only bolometers two rows away from bright
emission. We found small changes in $p$ ($\Delta p = -0.3$) and
$\theta$ ($\Delta\theta = -7^{\circ}$) but the differences between
the reductions are consistent with the errors derived from the
distributions in each map ($\sigma_p = 1.2,\,\sigma_\theta =
4^{\circ}$). As the first method uses bolometers which may have
jiggled onto polarised emission, we will use the second, more
conservative approach for the rest of the analysis. Our overall conclusions
are unaffected by this choice of sky-removal technique.

We have also compared our maps to the data presented by
\citet{matthews08} where the mean sky level was not removed from the
array (only the variations are removed) and bolometers showing excess
noise were flagged aggressively, resulting in the removal of around 30
per cent of the data. We have taken a slightly different approach,
flagging only the four bolometers with clear excess noise, plus one
that was dead: around 13 per cent of the data in total. The time
series of many bolometers removed by \citet{matthews08} displayed
momentary spikes but were otherwise in line with the rest of the
array. We de-spiked these by flagging any data more than $3\sigma$
from the mean. Overall, the mean $p$ and $\theta$ derived from both
reductions are consistent within the errors ($\Delta p = 1.0$ and
$\Delta\theta = -5^\circ$). We conclude that the data are robust to
minor changes in the data reduction procedure.

\subsubsection{Chopping}

We investigated the effects of chopping onto either polarised or
unpolarised emission, something which can potentially have a serious
impact on submm polarimetry data \citep{hildebrand00}. We synthesised
$Q$ and $U$ for our reference positions using either the radio values
scaled to 850\,\mic\ or assuming a value for $p$ and $\theta$ for the
interstellar medium (ISM) surrounding Cas~A (these are described in
more detail below). The $Q$ and $U$ values from our submm polarimetry
data were then corrected by adding half of the $Q$ and $U$ values
found at each reference position. The $I$ values were taken from the
850-\mic\ scan-maps, which comprised the \citetalias{dunne03}
flattened and unflattened reductions (hereafter A and B), the former
with large-scale undulations removed, and the map used by
\citet{krause04} (hereafter C). We created a simulated
`chopped' image from our measured $Q$ and $U$ values, and a `chopped'
$I$ image from the scan map after processing it with the chopping
simulator. This is intended to represent the data we measure. We then
correct for the chopping in $Q$, $U$ and $I$ and produce a `corrected'
map. This assumes, of course, that the scan-maps provide a good
representation of unchopped data (they are the best we have). Chopping
onto unpolarised flux should not affect $\theta$ but will influence
$p$. We have thereby simulated the range of plausible structures
around Cas~A to probe the possible systematics. In utilising the
unflattened scan-maps we have assumed that any structure is real and
not an artefact of the data reduction.

In the first instance we assumed that only flux associated with the
remnant was polarised and used the radio $Q$ and $U$ maps to estimate
the flux of polarised synchrotron in the reference beams. We accounted
for a rotation of 24$^\circ$ in going from 5-GHz to 850-\mic, in
line with the rotation measures derived by \citet{jones03} and
\citet{akr95}. We scaled the radio $Q$ and $U$ values to 850-\mic\
using the value of the synchrotron spectral index at the reference
points ($\alpha \sim -0.74$) and increased the polarised flux in the
reference positions by a factor three, roughly the average ratio of
$p_{\rm submm}/p_{\rm radio}$ that we observe, to allow for the reference
beams being polarised at the same level as the central field. The
results are summarised in Table~\ref{chopT}. We found that chopping
onto remnant emission polarised at the level seen in the submm would
have a small effect on the submm vectors, with an average increase in
$p$ of $+0.6$ per cent and an average rotation of $+0.4^{\circ}$.

\begin{figure*}
\includegraphics[width=8cm]{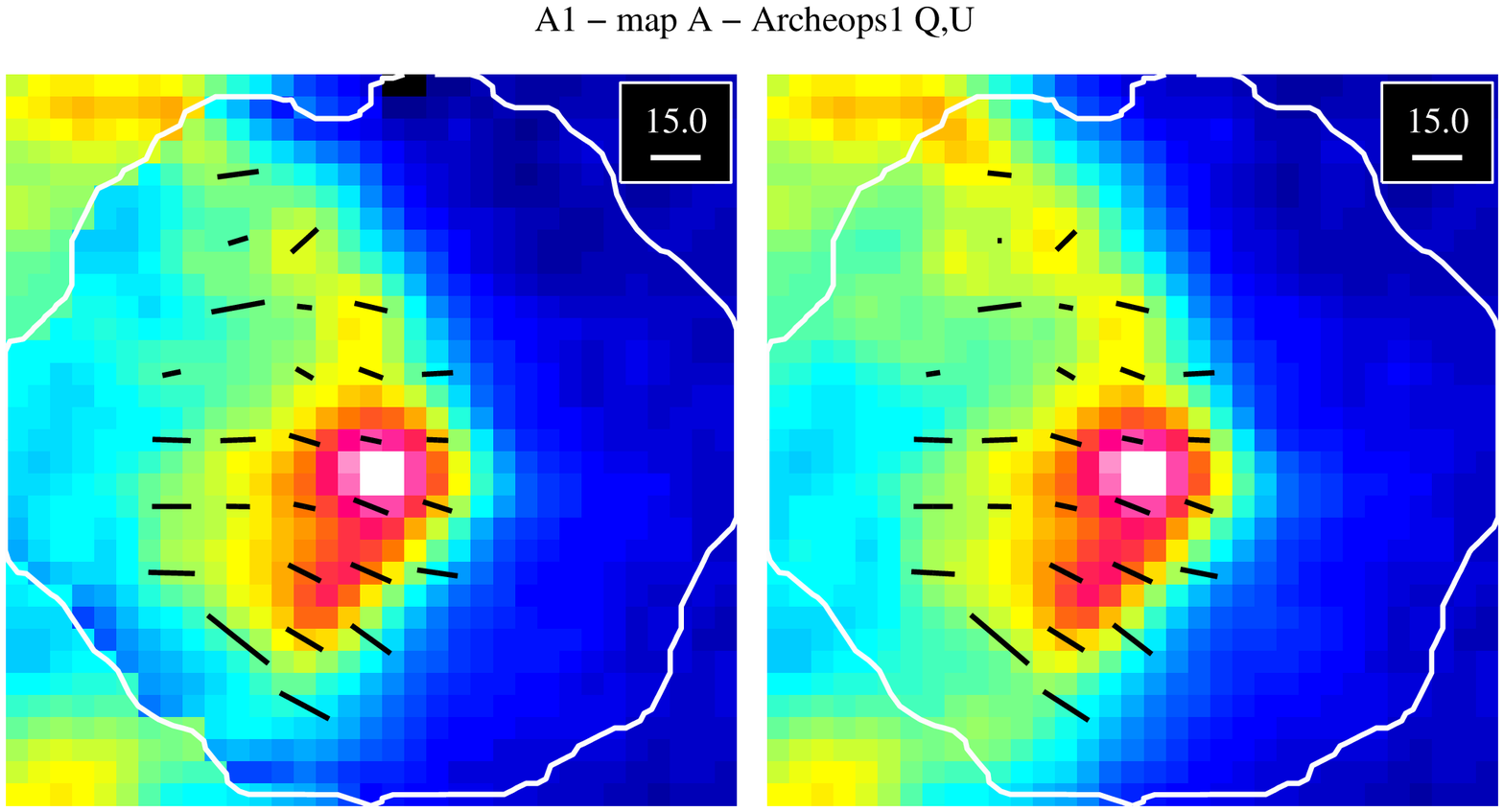}\hspace{10mm}
\includegraphics[width=8cm]{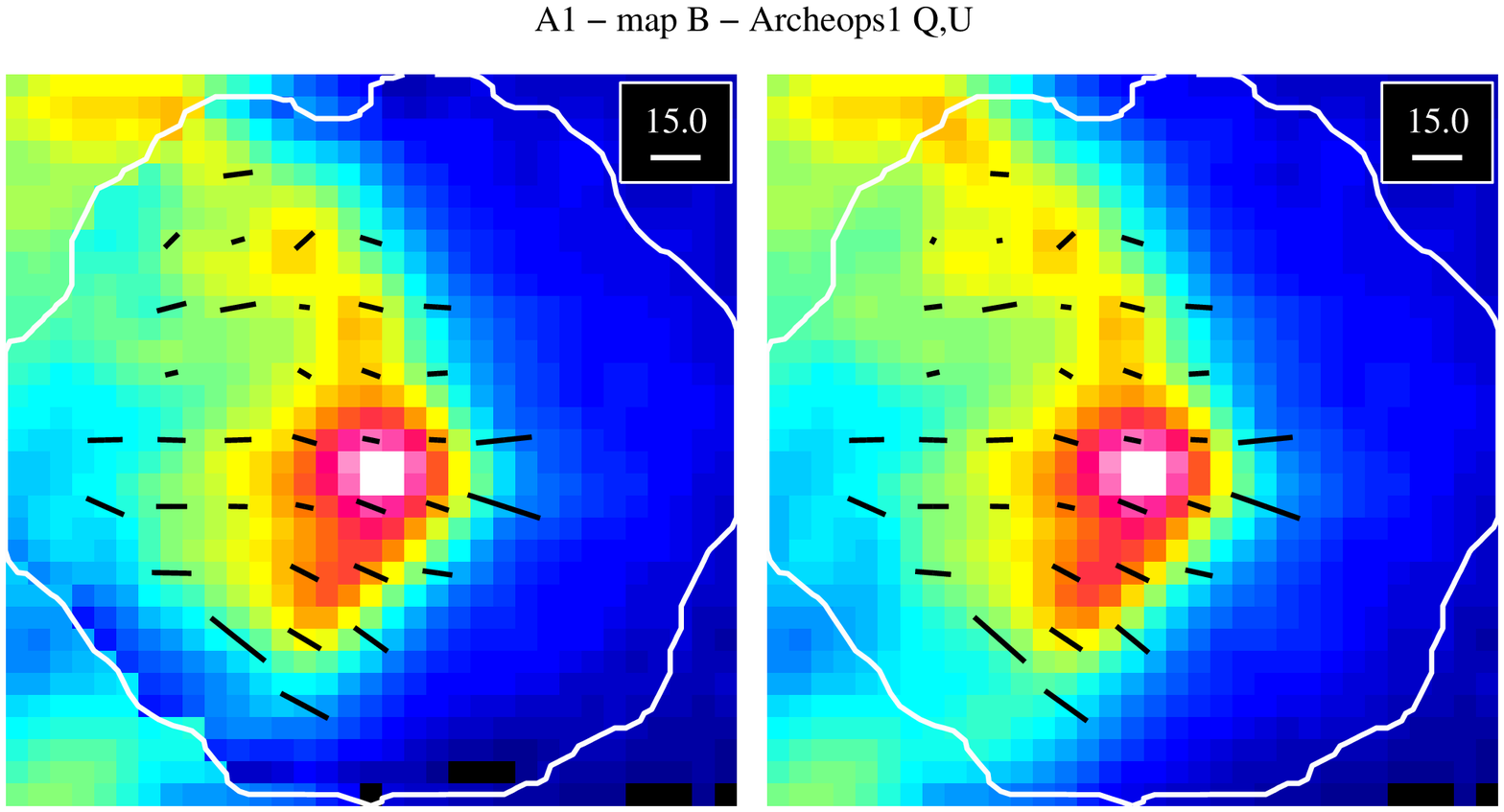}\\
\includegraphics[width=8cm]{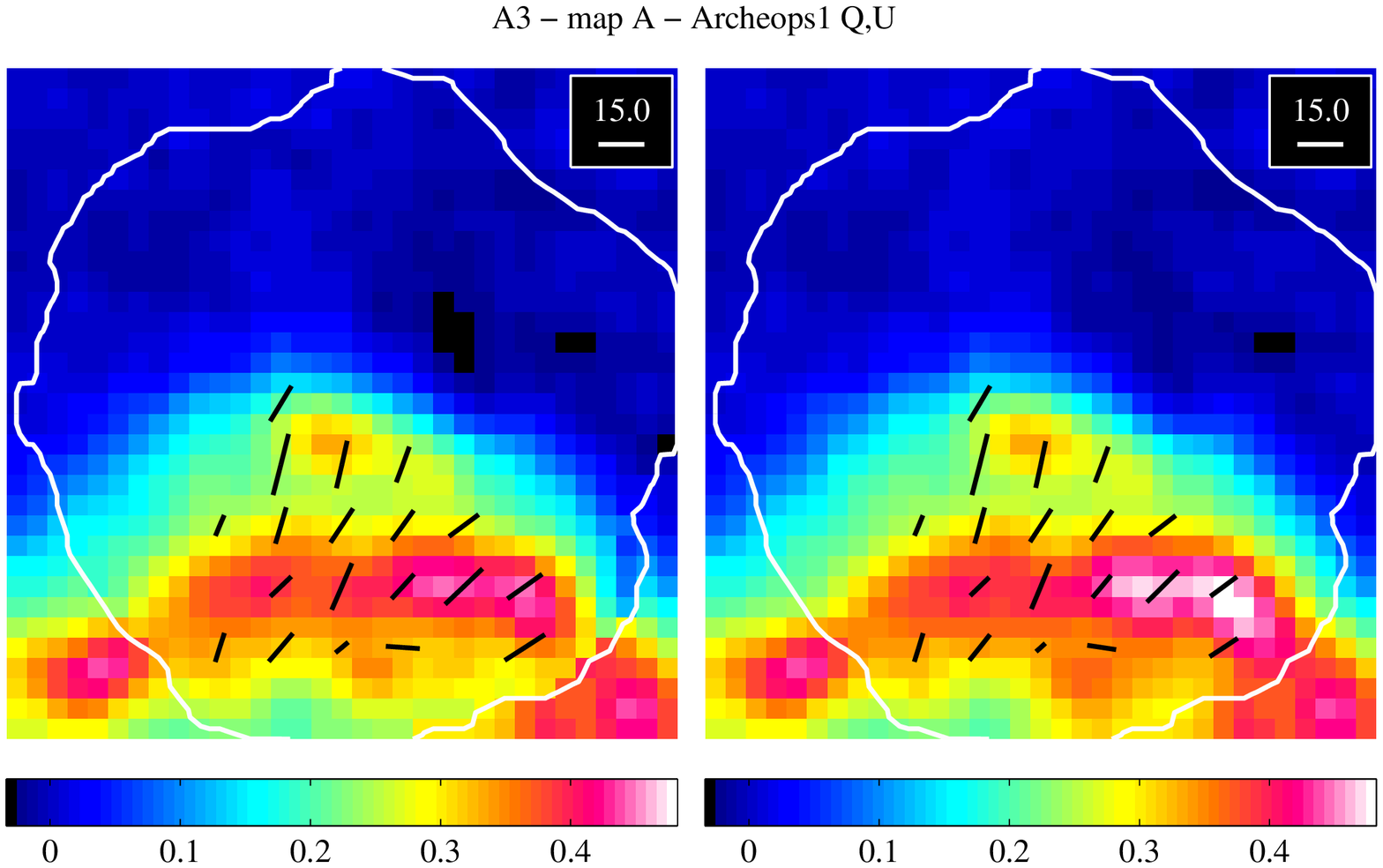}\hspace{10mm}
\includegraphics[width=8cm]{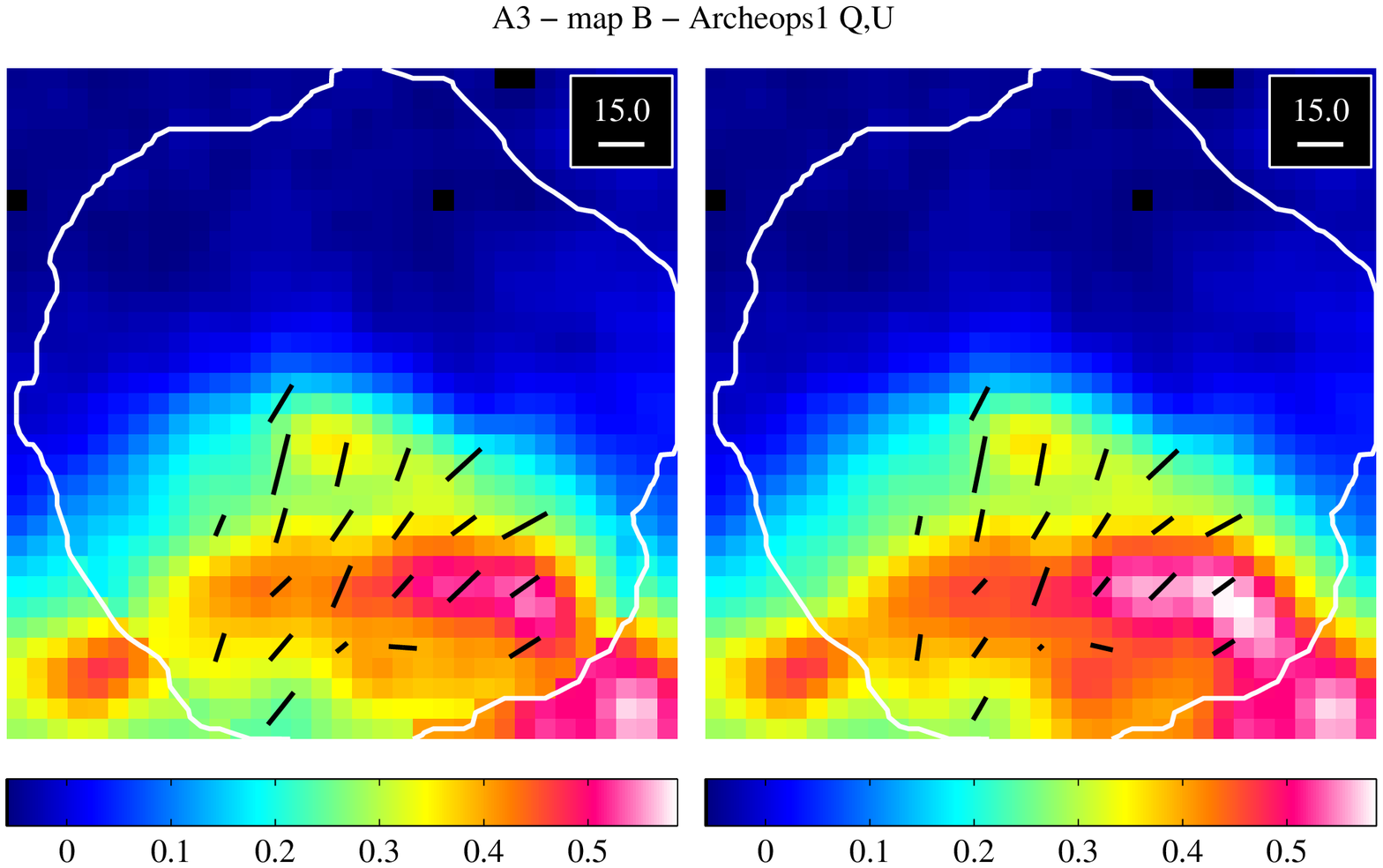}\\
\caption{Simulated chopped and unchopped images as
described in \S~\ref{sec:issues}. On the left of each sub-panel is the
`chopped' image, meant to represent the polarimetry data. Next to it
on the right is the corrected image. The upper row is for 
region W and the lower row is region N. The scan-maps used are A and B
and the $Q$ and $U$ model is {\em Archeops\/}~1
(see Table~\ref{chopT} and \S~\ref{sec:issues} for a detailed
description). The effects of chopping on $I$ can be mostly seen at the
edges of the fields where the edge of the remnant was chopped onto
(see Fig.~\ref{fig:regionsF}).}
\label{chopsimF}
\end{figure*}

We next investigated the possibility that any foreground submm
emission in the reference positions is polarised. We allow all of the
submm emission in the reference positions to be polarised at the level
found in clouds in the vicinity of Cas~A by the {\em Archeops\/} submm
balloon experiment \citep{benoit04}, which is 23 per cent at
$\theta=-\pi/3$ and 12 per cent at $\theta=\pi/8$. It is not certain
that the Cas~A foreground emission is polarised at this high level as
the clouds in \citeauthor{benoit04} are $>2^{\circ}$ away, but we
consider this as the most conservative case. We find that the
difference in the chopped and unchopped maps is $\Delta p =0.3-1.2$ in
field W and $\Delta p = 0.2-1.9$ in field N. The rotation
introduced by chopping is small, the largest being $+1.2^{\circ}$ in W
and $-2.7^{\circ}$ in N. These simulations are shown in
Fig.~\ref{chopsimF} and the results summarised in
Table~\ref{chopT}. None of the small changes introduced by chopping alter
our conclusions.

\begin{table*}
\caption{\label{chopT} Results of simulated chopping onto
polarised emission.}
\begin{tabular}{c|c|cccc|cccc}
\hline
\multicolumn{2}{c|}{}&\multicolumn{4}{c|}{W: western
field}&\multicolumn{4}{c}{N: northern field}\\
\multicolumn{1}{c|}{Model}&\multicolumn{1}{c|}{Submm $I$
map}&\multicolumn{1}{c}{$p_{\rm cor}$}&\multicolumn{1}{c}{$p_{\rm
chop}$}&\multicolumn{1}{c}{$\Delta_p$}&\multicolumn{1}{c|}{$\Delta\theta$}&\multicolumn{1}{c}{$p_{\rm
cor}$}&\multicolumn{1}{c}{$p_{\rm
chop}$}&\multicolumn{1}{c}{$\Delta_p$}&\multicolumn{1}{c}{$\Delta\theta$}\\
\hline
Scaled radio $Q,U \,\times 3$ &  A  & 11.1 &  11.9   & $+0.8$   & $+0.4$ &
13.0 & 13.7 & $+0.7$ & $ +0.9$ \\
                       &  B  & 8.2  &   8.7   & $+0.4$   & $+0.4$ &
11.1 & 12.2 & $+1.0$ & $+0.9$\\
                       &  C  & 10.6 &  11.1   & $+0.5$   & $+0.4$ &
See note$^{*}$ &   &    &   \\
\hline
{\em Archeops\/} 1 & A  & 10.7 & 11.9  & $+1.2$   & $+1.2$ &
12.5 & 13.7 & $+1.2$ & $-0.7$\\
$p=23$ per cent,\, $\theta=-\pi/3$ & B  & 8.2  & 8.7   & $+0.5$   & $+0.9$ &
10.3 & 12.2 & $+1.9$ & $-2.7$ \\
                          & C  & 10.4 & 11.1  & $+0.7$   & $+0.5$  &
&  &  \\
\hline
 {\em Archeops\/} 2& A & 11.6 & 11.9 & $+0.3$ & $ -0.3$  & 13.5 &
13.7 & $+0.2$ & $+0.5$\\
$p=12$ per cent, \, $\theta=\pi/8$ & B & 8.5  & 8.7  & $+0.2$ & $ -0.3$  & 11.7 &
12.2 & $+0.5$ & $+1.6$ \\
                        & C & 10.9 & 11.1 & $+0.2$ & $ -0.1$  &    &
 &    \\
\hline
Observed map          & polarimetry &   & $12.6\pm 1.2$  &      &    &  &
$15.1\pm 0.8$ & &   \\
\hline
\end{tabular}
\flushleft{\small {Notes: Column (1) refers to the model used for $Q$
and $U$ in the reference positions; (2) submm scan-map used to
determine the reference $I$ value and combined with the measured submm
$Q$ and $U$ to produce the quoted $p$ values (see text); (3) mean $p$
in the corrected map; (4) mean $p$ in the `chopped' map, error on the
mean $p$ is similar to that quoted for the polarimetry in the final
row of the table ($\sim 1$ per cent in each field); (5) $\Delta p
= p_{\rm chop} - p_{\rm cor}$; (6) $\Delta\theta = \theta_{\rm
chop}-\theta_{\rm cor}$. Columns 3--6 are repeated for positions W and
N which each have different reference chop positions. $^\ast$ scan-map
C cannot be used for position N as the background is too negative.}}
\end{table*}


\subsection{Submm and radio polarisation properties}
\label{submmradio}

Polarisation vectors were created as described in \S~\ref{submmobs} using
the same positions in the observed 850-\mic\ and 5-GHz images. We also
predict the observed polarisation expected at 850\,\mic, given the
measurements in hand at 5\,GHz and assuming that the excess submm
emission is unpolarised. We scale the 5-GHz $Q$ and $U$ fluxes using a
1.4\,GHz/5\,GHz spectral index map. We use the submm scan-map from
\citetalias{dunne03} as the $I$ image, since this exceeds the
predicted synchrotron flux by $\sim$30 per cent (due to dust emission,
either in the supernova remnant or in front of it). We then calculate
$p$ using the scaled radio $Q$ and $U$ data and the submm $I$
data. This is the polarisation we would expect if the submm $I$ emission
consisted of synchrotron and unpolarised dust.

\begin{figure}
\centering
\includegraphics[width=8.5cm]{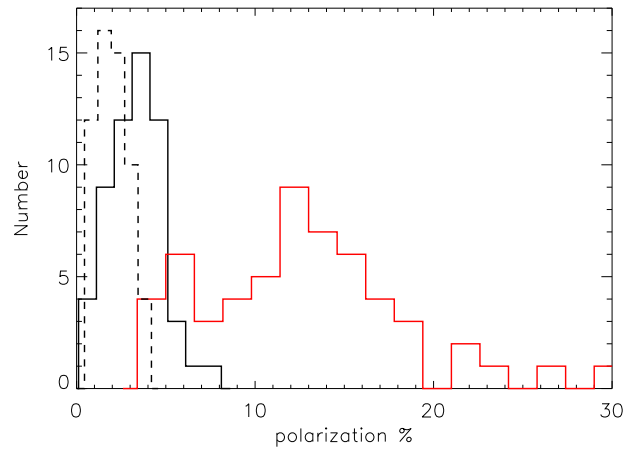}
\caption{Histogram showing the distribution of $p$ for the 5-GHz data
(black) and the 850-\mic\ data (red). The distributions are
significantly different. The predicted polarisation at
850\,\mic, using the 5-GHz data as described in the text, is shown by
the dashed line.}
\label{phistF}
\end{figure}

A comparison of the $p$ distributions in the radio and submm wavebands
is shown in Fig.~\ref{phistF} where is it clear that the submm
emission shows a significantly higher degree of polarisation than does
the radio. The means of the distributions are $p_{\rm submm} = 13.5\pm
0.8$ per cent and $p_{\rm radio}=3.7\pm 0.2$ per cent when selecting
vectors at $p/dp\geq 2$ and $dp<4$ per cent in the submm waveband. If
the SNR threshold is ignored, the mean $p_{\rm submm}$ drops slightly
to $12.9\pm 0.9$ per cent. The means are thus different at the
$>8\sigma$ level. If we compare to the predicted values at 850\,\mic\
(dashed histogram in Fig.~\ref{phistF}) the difference becomes even
more pronounced. The results of the chopping simulations suggest a
minor reduction in $p_{\rm submm}$ of order 0.7 per cent may be
required, but there appears to be no possibility of reconciling the
two distributions.

Having established a clear excess in polarisation at 850\,\mic, we now
wish to understand whether this difference is due to a reduction in
the level of synchrotron polarisation between 850\,\mic\ and 5\,GHz
via a depolarising mechanism in the remnant, or to an extra
source of polarised emission at 850\,\mic\ (i.e.\ aligned dust grains).

Synchrotron emission in supernova remnants should be highly polarised
($\sim$70 per cent) but observed polarisations at radio frequencies
in young remnants like Cas~A are typically much lower, implying that
the magnetic field is highly disordered on scales below an arcsec
\citep{milne87}. Another cause of reduced polarisation at radio
wavelengths is rotation due to the Faraday effect, in which the plane
of polarisation is rotated as the polarised waves pass through a
magnetised plasma. This effect is wavelength-dependent, with longer
wavelengths suffering more rotation. Emission from regions at
different depths in the plasma of Cas~A will be rotated by differing
amounts as they travel to the observer. This causes `internal
depolarisation' at the wavelengths which undergo the most rotation.

Depolarisation at radio frequencies has been studied in Cas~A and
found to be negligible at frequencies of $\gs$5\,GHz. From 5\,GHz to
2.2\,\mic, the fractional synchrotron polarisation remains roughly
constant \citep[4--7 per cent, ][]{jones03, akr95, kenney85} which
suggests that the much higher polarisations observed at 850\,\mic\
cannot be due to synchrotron radiation. In some specialised
geometries, depolarisation can lead to an increase in fractional
polarisation followed by a decrease at successively larger
wavelengths, but this occurs only under conditions of very strong
depolarisation \citep[$\ge 10$ --][]{cj80}, which does not apply in
Cas~A above 5\,GHz. The lack of depolarisation at 5\,GHz thus implies
that the high 850-\mic\ polarisations arise from an independent
source.

\begin{figure*}
\includegraphics[width=8.5cm]{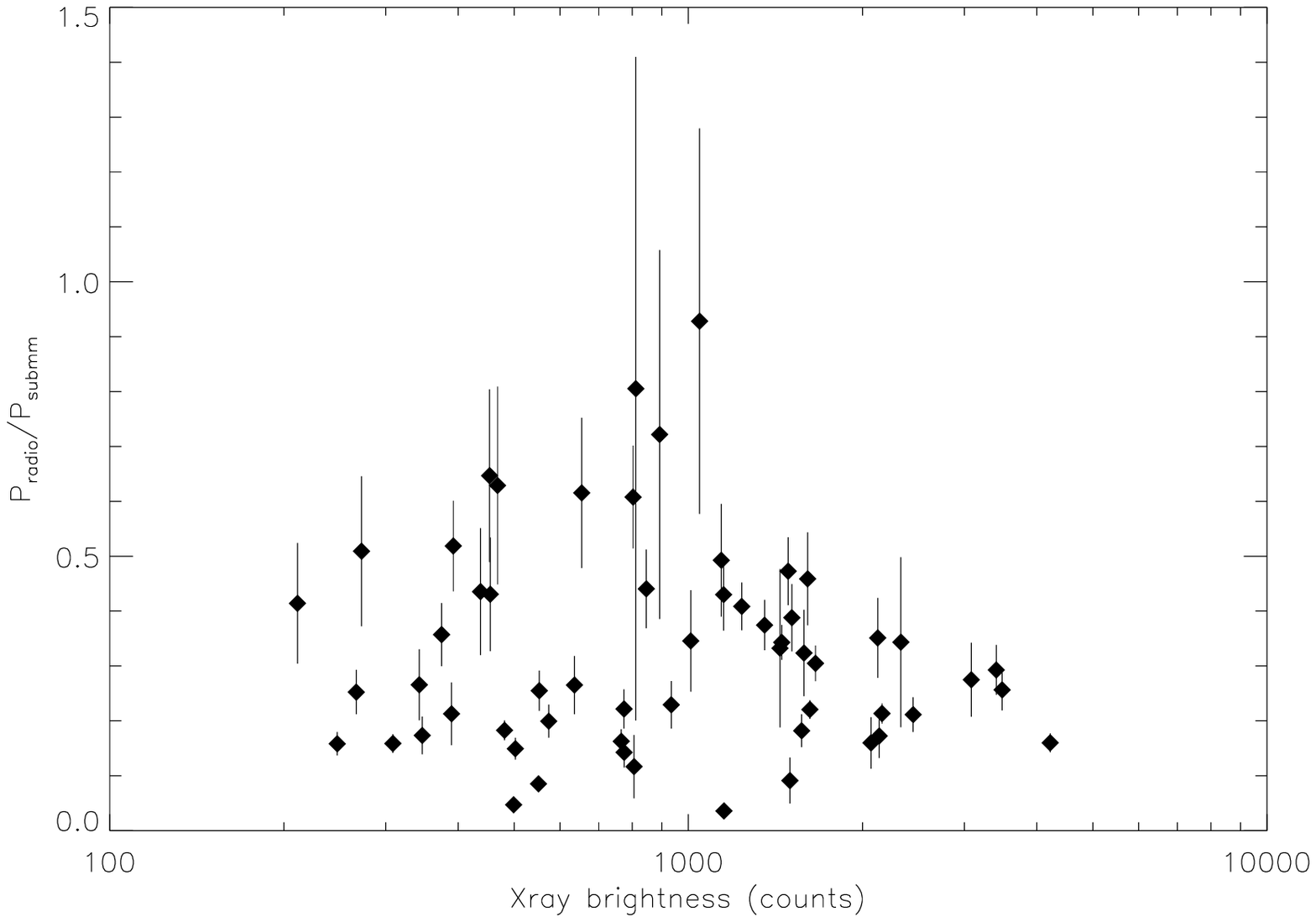}
\includegraphics[width=8.5cm]{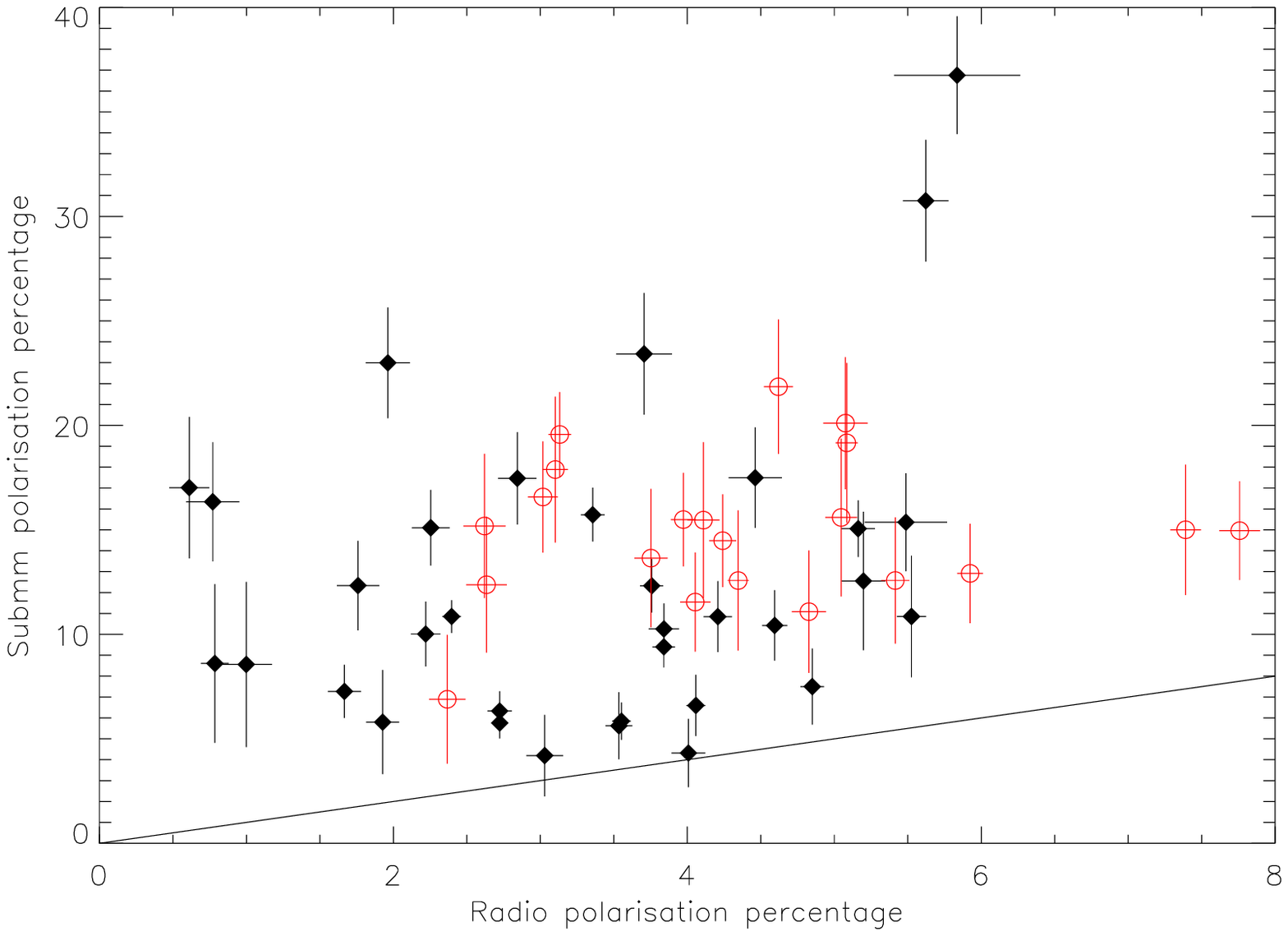}
\caption{\label{XraydepolF} {\em Left:} Depolarisation from 5\,GHz to
850\,\mic\ as a function of X-ray brightness. There is no clear
relationship, unlike the 1.4-GHz to 5-GHz depolarisation which
correlates strongly with X-ray brightness \citep{akr95}. {\em Right:}
Radio polarisation versus submm polarisation. The line shows a
one-to-one correspondence. Filled black diamonds are the Western
region while red open circles are the Northen region. There is no
obvious correlation.}
\end{figure*}

Another piece of evidence for internal depolarisation could have come
from comparisons of depolarisation with X-ray brightness, serving as a
proxy for the density of the depolarising thermal electrons.  Indeed,
\citet{akr95} showed a very strong correlation between the X-ray
brightness and the depolarisation of emission between 5\,GHz and
1.4\,GHz, arguing for internal depolarisation at the low radio
frequencies. However, no such dependence is seen in our comparison of
depolarisation between 850\,\mic\ and 5\,GHz and the X-ray brightness
(Fig.~\ref{XraydepolF}, left). Thus, the high fractional polarisations
at 850\,\micron\ must arise from another cause. This
850-\mic-independence is also demonstrated in the plot of $p_{\rm
radio}$ vs. $p_{\rm submm}$ in Fig.~\ref{XraydepolF} (right) where no
correlation is seen.

\begin{figure*}
\includegraphics[width=8.5cm]{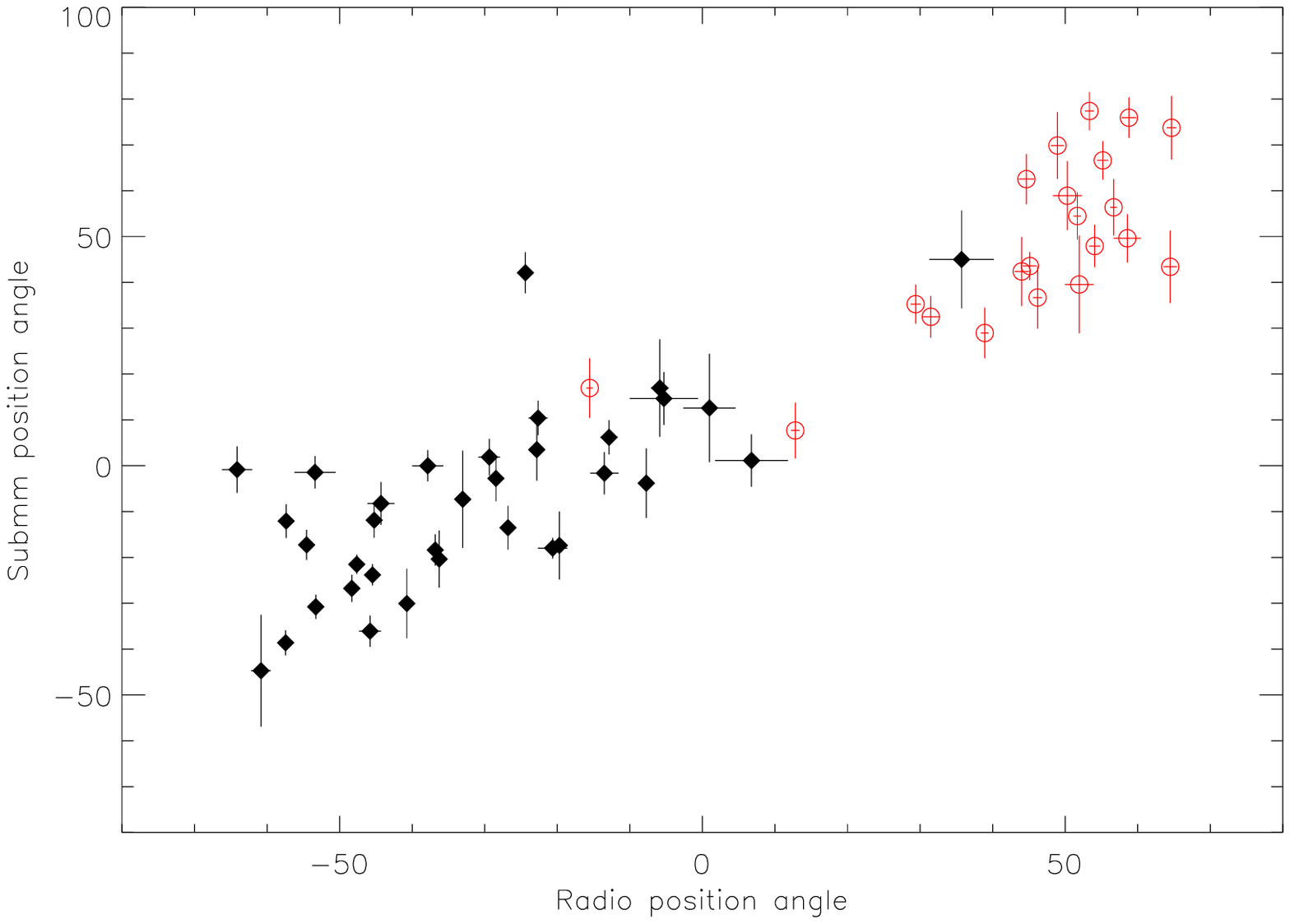}
\includegraphics[width=8.5cm]{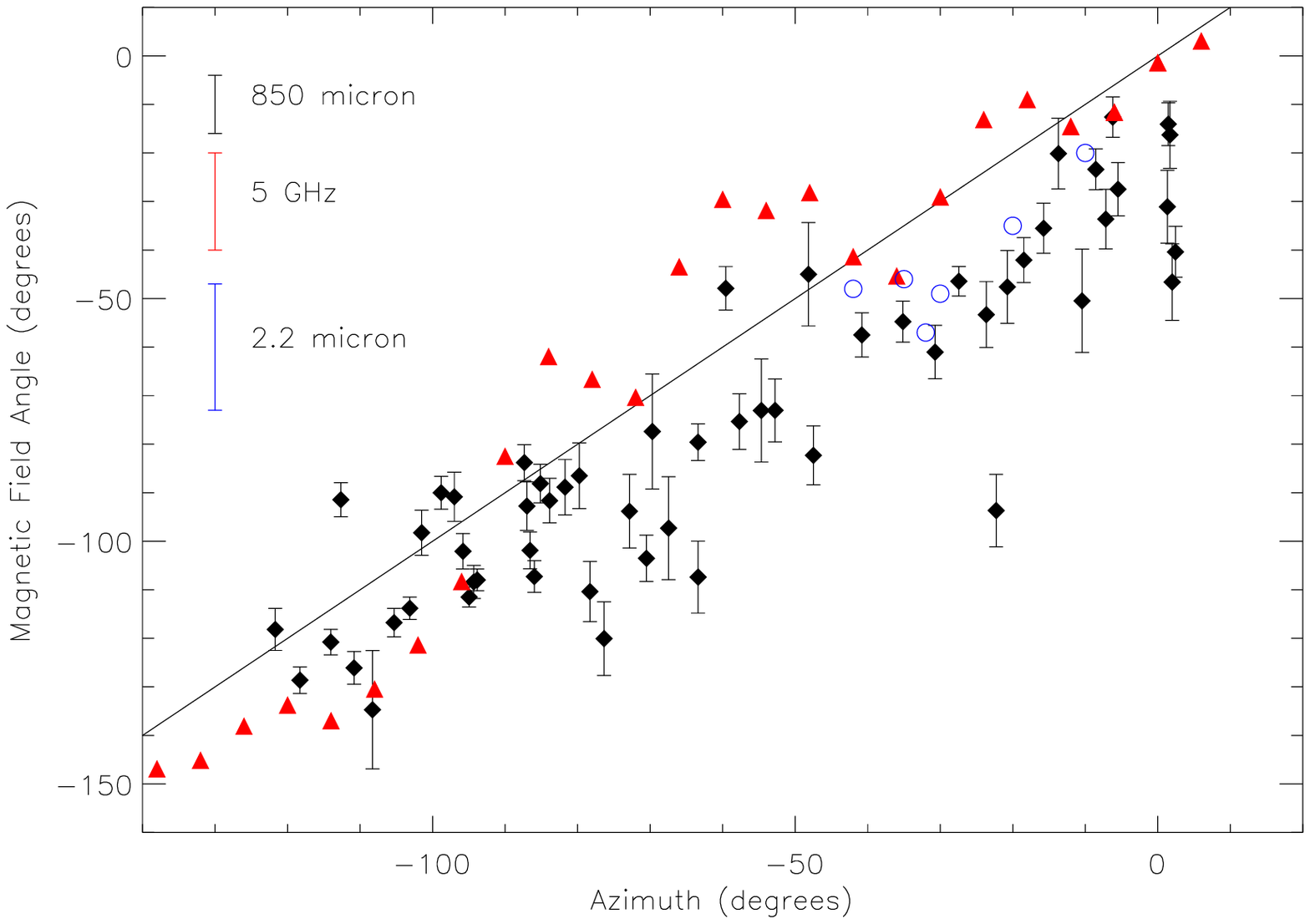}
\caption{\label{PAF} {\em Left\/}: Radio versus submm position
angles. Filled symbols represent the Western region while open symbols
represent the Northern region. There is a very good correlation
between the direction of the radio and submm polarisation vectors. {\em Right:\/} Inferred magnetic field angle (PA$-90^{\circ}$)
versus azimuth angle (0$^{\circ}$ for celestial north).  Solid
diamonds are 850-\mic\ PAs; red triangles are 5-GHz PAs from
\citet{akr95} corrected for the local RM as described in the text;
blue open circles are 2.2-\mic\ PAs from \citet{jones03}. Systematic
errors bars are shown in the top left. The solid line shows a pure
radial alignment. }
\end{figure*}

We now turn to the PA of the polarisation. In
Fig.~\ref{PAF}(left) we show the PAs ($\theta$) of the radio versus
the submm. Filled symbols indicate the Western region and open symbols
the Northern region. There is a very good correlation between PAs
contrary to the lack of correlation between polarisation
strengths. This suggests that the submm polarisation is largely
arising from a different physical mechanism than the synchrotron radio
polarisation. However, both polarisation signals reflect the same
underlying magnetic field in Cas~A. This is further demonstrated
in Fig.~\ref{PAF}(right) where we shows the PAs of the radio and submm
and 2.2-\mic\ polarisation vectors \citep{jones03} as a function of
azimuth. The radio measurements have been corrected for the local
Faraday rotation, redetermined from the data of
\citet{akr95}. They quoted an average value for the rotation measure
(RM) of $-110\,\rm rad\,
m^{-2}$, while we determine a range of values from $-50$ to $-302\,\rm
rad\,m^{-2}$ for our range of azimuths. The extrapolated rotations at
850\,\mic\ are $<$0.01 degrees, so no correction for Faraday rotation
has been made for those, or for the 2.2-\mic\ data.

There is a very good correlation between all three sets of
measurements, suggesting that whatever is responsible for the extra
polarisation signal in the submm waveband is related to the same
magnetic fields that are responsible for the synchrotron emission at
5\,GHz and 2.2\,\mic\, and, perforce, is intrinsic to Cas~A and is not
due to foreground material. The small mean offset between the
corrected 5-GHz PAs and those at 850\,\mic\ and 2.2\,\mic\ are
consistent with the systematic uncertainty in the RM correction.  To
first order the PAs all provide strong evidence for a radial magnetic
field, as is common in young supernova remnants \citep{akr95}.

If we take the submm and 2.2-\mic\ magnetic field PAs at their face
value, ignoring the systematic uncertainties, they agree very well
with each other and are systematically offset from a radial angle by
$\sim$10-20$^o$ in a clockwise direction. At present, there is no
apparent reason for this offset.

\section{Discussion}
\label{dust}

Since we cannot explain the submm polarisation as arising from
synchrotron emission we will now consider the most likely alternative
-- that it is due to dust aligned with the magnetic field in the
remnant. We will estimate $p_{\rm d}$, the polarisation fraction of the
dust in the remnant. We know that the submm flux we measure is a
combination of radio synchrotron and thermal emission from dust, such
that $I_{\rm s} = I_{\rm r} + I_{\rm d}$. Similarly we can write that
$Q_{\rm s} = Q_{\rm r} + Q_{\rm d}$ and $U_{\rm s} = U_{\rm r} +
U_{\rm d}$. We can also express $Q$ and $U$ in terms of $I$, $p$ and
$\theta$, as $Q=Ip\cos(2\theta)$ and $U=Ip\sin(2\theta)$. Combining
these equations and solving for dust polarisation, $p_{\rm d}$, gives:

\begin{equation}
\label{RM}
p_{\rm d} = \frac{\left[(I_{\rm s}p_{\rm s} -
I_{\rm r}p_{\rm r})^2 + 2\, I_{\rm s}p_{\rm s}I_{\rm
r}p_{\rm r}(1-\cos 2(\theta_{\rm s} -
\theta_{\rm r}))\right]^{1/2}}{I_{\rm d}}
\end{equation}

\noindent
which for $\theta_{\rm s} = \theta_{\rm r}$ simplifies to 

\begin{equation}
\label{dp0}
p_{\rm d} = (p_{\rm s} - p_{\rm r})\frac{I_{\rm r}}{I_{\rm
d}} + p_{\rm s}
\end{equation}

\begin{figure}
\includegraphics[width=8.5cm]{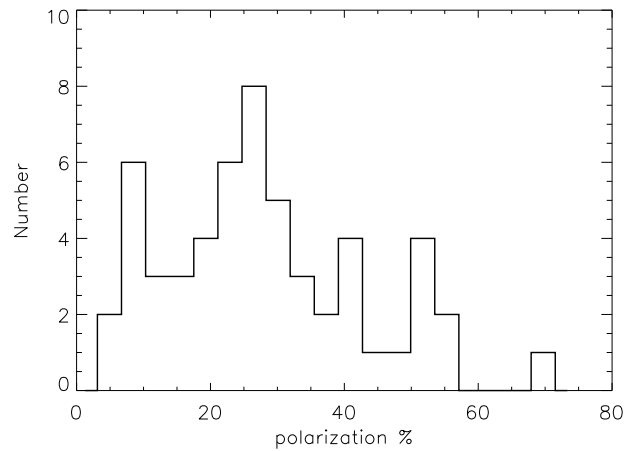}
\caption{\label{phistdF} Histogram of dust polarisation fraction,
$p_{\rm d}$, calculated using Eqn.~\ref{dp0}.}
\end{figure}

The fractional dust polarisation calculated using Eqn.~\ref{dp0} is
unprecedented, with an average $p_{\rm d}= 30\pm 2$ per cent. A
histogram of $p_{\rm d}$ is shown in
Fig.~\ref{phistdF}. Eqn.~\ref{dp0} assumes that the polarised
synchrotron and dust radiation is emitted with the same intrinsic PA
(i.e.\ aligns with the $B$ field in the same way) and that any
rotation between 5-GHz and 850-\mic\ vectors is due to Faraday
rotation at 5\,GHz. This is a reasonable assumption given
Fig.~\ref{PAF}. We obtain a slightly higher value for $p_{\rm d}$ (33
per cent) if we use the difference between the RM corrected 5-GHz PA
and the submm PA in Eqn.~\ref{RM} instead.

\begin{figure}
\includegraphics[width=8.5cm]{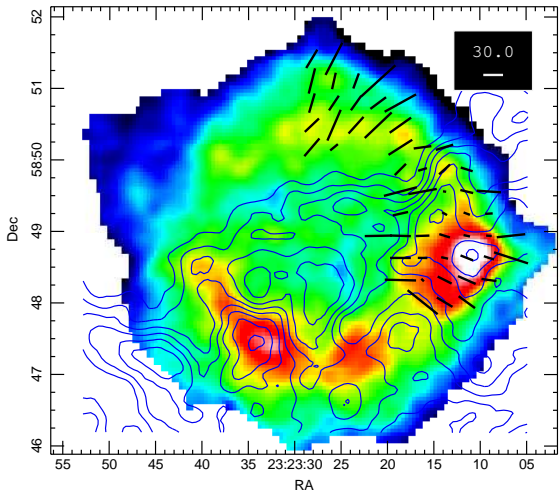}
\caption{\label{dustpolF} Synchrotron-subtracted 850-\mic\ image,
showing the location of the dust. Overlaid are the dust polarisation
vectors calculated using Eqn.~\ref{dp0}. Also shown in blue are
CO(2--1) contours, at the same resolution \citep[][(in
preparation)]{eales}. CO contours coincide with the western dust peak,
but the dust polarisation at these positions indicates that some of
the 850-\mic\ emission must be due to dust in the remnant. The
decrease in dust polarisation at the CO peak in the west is consistent
with some fraction of the submm emission at this location being due to
foreground material, associated with the CO. Note that there is no CO
emission coincident with the dust emission in the north.}
\end{figure}

We can create a map of 850-\mic\ dust flux density, $I_{\rm d}$, by
scaling the 5-GHz map to 850\,\mic\ as described in
\S~\ref{submmradio}. Subtracting the scaled synchrotron map from the
850-\mic\ image leaves the 850-\mic\ emission due to dust. This dust
map\footnote{This map differs slightly from that in
\citetalias{dunne03} as we have used a more accurate, spatially
varying spectral index to extrapolate the radio synchrotron flux to
the submm waveband.} is shown in Fig.~\ref{dustpolF}, with the dust
polarisation calculated from Eqn.~\ref{dp0}. Over-plotted are CO(2--1)
contours from \citeauthor{eales} (in preparation). This shows three
things: first, there is dust emission in the northern part of the
remnant, but no corresponding CO emission; second, the level of submm
polarisation anti-correlates with the CO contours; the dust
polarisation being lowest at the western peak, consistent with the
suggestion by \citet{krause04} and \citet{wb05} that its submm
emission is contaminated by foreground clouds; third, there is
polarised dust emission in both the north and the west.

The polarisation signal is correlated with the magnetic field in the
remnant, which means that some of the emission in the western region
must be associated with the remnant. None of dust intensity or
polarised flux in the north is associated with molecular
material. Thus the finding of \citet{krause04} that there is `no cold
dust in Cas~A' is demonstrably incorrect. The average dust
polarisation around the western peak is $13.7\pm 1.4$ per cent; in the
north, away from the CO, it is $39.4\pm 3.2$ per cent. If we assume
that the intrinsic value of $p_{\rm d}$ is similar throughout the
remnant, we can use the uncontaminated values in the north to estimate
the contribution of the foreground material to the western peak. Since
$p \propto 1/I$, we simply correct the dust flux in the western peak
by the ratio $13.7/39.4 = 0.34$, which gives a flux density of $\sim
0.8\,$Jy intrinsic to Cas A at the Western peak. In order to make
a comprehensive correction for the intervening foreground material a
full set of molecular gas tracers must be observed over the whole
remnant. This data set does not yet exist and therefore we cannot make
a definitive measurement of the total submm flux associated with the
remnant at this time.  We can, however, estimate a conservative lower
limit by assuming that only the polarised flux density at 850\,\mic
($\sim 30$ percent) is intrinsic to the remnant. The integrated flux
density from the dust map in Fig.~\ref{dustpolF} is $S_{850}=
20.1$\,Jy\footnote{This is higher than the value of 15.9 Jy quoted in
\citetalias{dunne03} due to the lower synchrotron contribution at
850\,\mic\ when using a varying radio spectral index}, and taking the
average 30 per cent polarised fraction as the minimum gives us a total
lower limit on the dust flux for Cas~A at 850\,\mic\ of 6.0\,Jy, based
on our submm polarimetry.\/

To convert this flux density into a dust mass we must assume a value
for the dust mass absorption coefficient $\kappa_{850}$. In our
previous work \citepalias{dunne03} we used a value which was derived
from laboratory studies of cosmic dust analogues which were amorphous,
non-spherical or aggregates. This gave a high value of
$\kappa_{850}=0.76\,\rm m^2\,kg^{-1}$. If we apply this value to our
flux measurement and use a temperature of $T_d=20$ K (which comes from
fitting the SED from the IR-submm) we get a dust mass estimate for
Cas~A of $\sim 1.0$\,\Msun. This value is consistent with theoretical
predictions for supernova dust yields
\citep[e.g.][]{tf01,nozawa03,bs07} and sufficient to explain the dust
at high redshift \citep{me03, dwek07}. This is also an upper limit to
the possible mass of dust in Cas~A given the masses of condensible
elements formed in supernovae with progenitor masses in the range
applicable to Cas~A (13-20 \Msun -- as suggested by the recent
determination of the type of Cas~A as a IIb by
\citet{krause08}). According to nucleosynthetic models by
\citet{lc03}, supernovae with progenitors in this mass range could
produce 0.5-1.0 \Msun of dust if the condensation efficiency
were 100 per cent. 

We can also look at new values for $\kappa_{850}$ which have been
derived from models of dust formation in supernovae similar to
Cas~A. Values range from $\kappa_{850}=0.2\,\rm m^2\,kg^{-1}$ from
\citet{bs07} to $\kappa_{850} = 0.049 \,\rm m^2\, kg^{-1}$ for a
Cas~A-like case from \citep[][in preparation]{nozawa03,
kozasa}. These $\kappa$ values produce dust masses in the range
3.8--15\, \Msun. Given the above discussion on the available mass of
condensible elements, these values are not physically plausible. This
suggests that the dust which is emitting the polarised radiation at
submm wavelengths is not currently predicted by the dust formation
models and is not necessarily the same as the dust which is emitting
at the shorter FIR wavelengths seen by {\em Spitzer\/} and {\em
IRAS\/}. The supernovae formation models currently only deal with
spherical grains, but in order to produce polarised emission the
grains must be elongated to some degree. Further development of the
dust formation models may be required in order to produce grains which
can reproduce the quantity and polarised nature of the submm emission
in Cas~A. In order to have a physically plausible mass of dust the
emissivity of the dust must be high - at least as high as that found
in the laboratory studies of amorphous and aggregate grains.\/

The highly elongated iron needle grains proposed by
\citeauthor{dwek04} alleviate the emissivity issue, as for
long enough axial ratios they have emissivities many orders of
magnitude higher than spheroidal non-conducting grains. Graphite
whiskers may also provide a similar solution, especially in light of
their recent discovery in meteorites \citep{fries08}. Iron or graphite
needles could reduce the dust mass required although a full
analysis of the properties of such grains which match the new
observations is beyond the scope of this paper. We therefore do not
rule out such exotic grains and will consider them in more detail in a
future paper \citep[][in preparation]{gomez}.\/

\subsection{Grain alignment}

What does this polarisation fraction tell us about the properties of
the dust in Cas~A? Average polarisation fractions in typical
interstellar and molecular clouds are of the order 2--7 per cent
\citep{hildebrand00, matthews02, cc07} though some clouds have values
as high as 15--20 per cent \citep{benoit04}. We are unaware of any
measurements as high as those quoted here, which suggests that either
the dust or the alignment mechanisms in Cas~A differ - perhaps
unsurprisingly -- from those in the general ISM.

Observations of polarised starlight and thermal emission from dust
mean that there must be a population of dust grains which are
non-spheroidal and a mechanism capable of aligning the grains so that
an appreciable polarisation signal can be detected. The details of how
interstellar dust aligns so efficiently, despite impacts by gas atoms
acting to randomise their orientations, has kept theorists busy for
half a century. The original proposal by \citet{dg50, dg51} suggested
paramagnetic alignment, where the axis of rotation of the slightly
elongated grains becomes aligned with the $B$ field though magnetic
dissipation. This theory was unable to explain the polarisation levels
observed in the ISM and star-forming regions because the timescales
for alignment are longer than the timescales for randomisation by
collisions with the gas.

There have been many significant improvements to the theory of grain
alignment in recent years -- see \citet{lazarian07} for an excellent
review. Several mechanisms are now thought to operate in different
astrophysical environments. Briefly these comprise radiative torques
\citep{dolginov72, dw96, lh07b, hl08a}, mechanical alignment in a
supersonic or subsonic gas flow \citep{gold52, rhm95, lazarian97,
yl03, lh07a} and super-paramagnetic alignment \citep{lh08}.

The strong and turbulent magnetic field in Cas~A \citep[$B\sim
0.5$\,mG --][]{wright99}, together with the large abundances of heavy
elements and the presence of a hot X-ray plasma, means that certain
mechanisms may be particularly applicable to the environment of
Cas~A. Pinwheel torques arising from the interaction of grains and
electrons in a hot plasma \citep{hl08b}, supersonic mechanical
alignment, sub-sonic mechanical alignment through MHD turbulence
\citep{yl03} or grains with super-paramagnetic (e.g.\ Fe) inclusions
\citep{lh08} can all lead to highly efficient alignment. The
timescales they have to operate is small in astrophysical terms, as
the explosion of Cas~A occurred only $\sim$300 years ago. We note that
the low synchrotron polarised fraction indicates a turbulent and
disordered magnetic field on sub-arcsecond angular scales. This acts
to reduce the observed polarisation as random orientations are
averaged within the synthesised beam. If the dust grains were aligned
with this tangled small-scale magnetic field then we might expect a
similar beam depolarisation effect to operate at submm
wavelengths. However, this would give rise to unphysical intrinsic
dust polarisation fractions and so we believe that the alignment
mechanism must be operating on larger scales. The radial $B$ field in
Cas~A is ordered on large scales and thought to arise from
Raleigh-Taylor instabilities at the reverse shock boundary
\citep{gull75}. These large-scale radial motions may also be
responsible for aligning the dust in Cas~A.

Even if full alignment is theoretically possible, the line of sight
angle to the B-field in Cas~A together with the random alignment of
orientations along the line of sight and within the beam should all
act to reduce the measured polarisation. That we still observe such a
high $f_{pol}$ indicates that the grains themselves are emitting
strongly polarised radiation. For ideal alignment and line-of-sight
viewing conditions $f_{pol} \sim 30$ percent can be acheived with only
moderately non-spherical grains of astronomical silicate, with axial
ratios of between 1.4 -- 1.7 for oblate and prolate grains
respectively \citep{padoan01}. Thus the level of polarisation {\em per
se\/} is not, on its own, strong evidence for needle like grains.
An estimate of the aligned fraction would be required in order to
constrain the shape of the grains.\/

A full application of the various grain alignment models to Cas~A is beyond
the scope of this paper but these data should provide an interesting
test of aligment theories and in time should also shed light on the
nature of the grains responsible for the polarised submm emission.

\section*{Summary}

\begin{itemize}
\item
We have discovered unprecedented levels of submm polarisation towards
the Cas~A supernova remnant, significantly in excess of that expected
from the radio synchrotron, and correlated with the magnetic field
direction in Cas~A.
\item
There is no currently known way to produce the polarised submm
emission from a synchrotron process; a depolarising mechanism capable
of producing a peak in $f_{\rm pol}$ as a function of wavelength would
be required.
\item
The polarised emission is therefore strong evidence that a significant
fraction of the cold dust detected by \citetalias{dunne03} is
associated with the supernova remnant. Further strong evidence is that
there is both polarised and unpolarised dust emission in the northern
region which is not affected by foreground molecular clouds.
\item
A high dust emissivity is required in order to stay within the
constraints of the mass of condensible elements available to form
dust. This limit on the emissivity is not consistent with that
currently predicted for dust formed in supernovae.\/
\item
The strength of the polarisation signal is unprecedented in the
general ISM and indicates a highly efficient alignment mechanism at
work in Cas~A. These data provide a stringent test of grain alignment
mechanisms, given the very short timescale available, $\sim$300\,yr.
\item
Alternatively, the strong signal may arise in exotic grains such as
iron or graphite needles. Such grains would naturally produce a higher
emissivity and keep within the mass budget.
\item
Higher-resolution polarimetry at 850\,\mic\ is required to confirm the
polarisation signal and provide a comparison at the spatial resolution
of the radio synchrotron emission. Measurements of the high-frequency
(100--200-GHz) synchrotron polarisation are required to test whether a
perverse radio depolarisation mechanism is at work.
\item
Future far-IR and submm polarimeters, such as those planned for
SCUBA-2 \citep{bastien05} and {\em FIRI} \citep{hi08}, will be able to
image the submm polarisation in Cas~A at higher resolutions and
frequencies.
\end{itemize}

\section*{Acknowledgements}

We wish to thank the JCMT staff that were involved in commissioning
and maintaining the SCUBA Polarimeter. The data presented here were
awarded under programme M04AU10. We also thank A.\ Chrysostomou, R.\
Curran, J.\ Greaves and G.\ Scheiven for helpful discussions on
polarimetry data reduction, T.\ J.\ Jones for comments on the 2.2-\mic\
data, S.\ Bianchi and T.\ Nozawa for discussions related to the
emissivity, A.\ Lazarian and B.\ Draine for very helpful comments
on grain alignment and polarisation. 

\bibliographystyle{mnras}
\bibliography{CasApol_proofs}

\bsp

\end{document}